\documentclass[RNAAS]{aastex631}
\usepackage{xcolor}

\begin{document}

\title{Comparing the Photometric Calibration of DESI Imaging and Gaia Synthetic Photometry}
\correspondingauthor{Rongpu Zhou}
\email{rongpuzhou@lbl.gov}

\author[0000-0001-5381-4372]{Rongpu Zhou}
\affiliation{Lawrence Berkeley National Laboratory, 1 Cyclotron Road, Berkeley, CA 94720, USA}

\author[0000-0002-4928-4003]{Arjun Dey}
\affil{NSF's NOIRLab, 950 N. Cherry Ave., Tucson, AZ 85719, USA}

\author[0000-0002-1172-0754]{Dustin Lang}
\affiliation{Perimeter Institute for Theoretical Physics, 31 Caroline Street North, Waterloo, ON N2L 2Y5, Canada}
\affiliation{Department of Physics and Astronomy, University of Waterloo, Waterloo, ON N2L 3G1, Canada}

\author[0000-0002-2733-4559]{John Moustakas}
\affiliation{Department of Physics and Astronomy, Siena College, 515 Loudon Road, Loudonville, NY 12211, USA}

\author[0000-0002-3569-7421]{Edward F.~Schlafly}
\affiliation{Space Telescope Science Institute, 3700 San Martin Drive, Baltimore, MD 21218, USA}

\author{David J.~Schlegel}
\affiliation{Lawrence Berkeley National Laboratory, 1 Cyclotron Road, Berkeley, CA 94720, USA}

\begin{abstract}
The relative photometric calibration errors in the DESI Legacy Imaging Surveys (LS), which are used for DESI target selection, can leave imprints on the DESI target densities and bias the resulting cosmological measurements. We characterize the LS calibration systematics by comparing the LS stellar photometry with Gaia DR3 synthetic photometry. We find the stellar photometry of LS DR9 and Gaia has an \textsc{rms} difference of 4.7, 3.7, 4.4 mmag in DECam $grz$ bands, respectively, when averaged over an angular scale of $27\arcmin$.
There are distinct spatial patterns in the photometric offset resembling the Gaia scan patterns (most notably in the synthesized $g$-band) which indicate systematics in the Gaia spectrophotometry, as well as honeycomb patterns due to LS calibration systematics. We also find large and smoothly varying photometric offsets at $\mathrm{Dec}<-29.25\degr$ in LS DR9 which are fixed in DR10.
\end{abstract}


\section{Introduction} \label{sec:intro}

We assess the relative calibration of the DESI Legacy Imaging Surveys \citep{dey_overview_2019, zou_project_2017, schlegel_dr9} by comparing with Gaia DR3 synthetic photometry \citep{gaia_xp_spectra,gaia_xp_de_angeli}. We 
analyze two LS data releases (DR9 and DR10). Here, we present the maps for DR9; additional figures and data for both DR9 and DR10 are included on zenodo\footnote{\url{https://doi.org/10.5281/zenodo.7884447}}. 

\section{Data} \label{sec:data}
To synthesize Gaia photometry in the DECam passbands, we 
use the Gaia XP Continuous Mean Spectra\footnote{\url{http://cdn.gea.esac.esa.int/Gaia/gdr3/Spectroscopy/xp_continuous_mean_spectrum/} (version 1.2.3)} 
and the GaiaXPy\footnote{\url{https://gaia-dpci.github.io/GaiaXPy-website/}} packages. Gaia spectra are only available for stars with $G\le17.65$, and there are significant spatial gaps in the coverage (see the additional figures on zenodo). We cross-match Gaia spectroscopic sources to LS sources with a search radius of $0.1^{\prime\prime}$. We only include stars with $0.6<\mathrm{BP}-\mathrm{RP}<2.5$, $G\ge13.5$, and with valid flux measurements in $g$-, $r$- and $z$-bands. We remove blended sources by requiring $\mathrm{FRACFLUX}<0.1$ in $grz$\footnote{\url{https://www.legacysurvey.org/dr9/files/\#sweep-brickmin-brickmax-fits}.}.
We remove saturated sources in LS in each filter by using ANYMASK=0\footnote{\url{https://www.legacysurvey.org/dr9/bitmasks/\#allmask-x-anymask-x}.}. The saturation limit is fainter for Decl.$\ge32^\circ$ (especially in $z$-band), and as a result, we have fewer (and typically fainter) stars available in the north. 

The Gaia-DECam synthetic photometry is not ``standardized'' (see \S2.2 of \citealt{gaia_xp_spectra}) compared to the observed LS photometry, and it has systematic offsets (of up to 8 mmag) which depend on ($BP-RP$) and $G$. We correct for these offsets by fitting them with polynomials of $BP-RP$ and $G$ separately for the northern and southern LS. 
We obtain the polynomial coefficients using stars in $-10^{\circ}<\mathrm{Dec}<+10^{\circ}$ for DR9 South and DR10 and in all of DR9 North. In the regions where the fits are performed, the average photometric offset is zero by definition, i.e., we do not measure the \textit{absolute} calibration offset. 

\section{Results} \label{sec:results}

\begin{figure}
    \centering
    \resizebox{0.6\columnwidth}{!}{\includegraphics{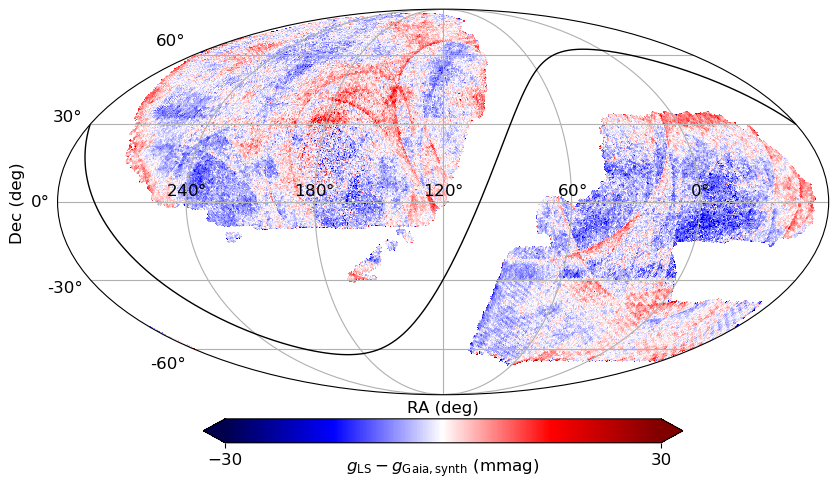}}
    \resizebox{0.6\columnwidth}{!}{\includegraphics{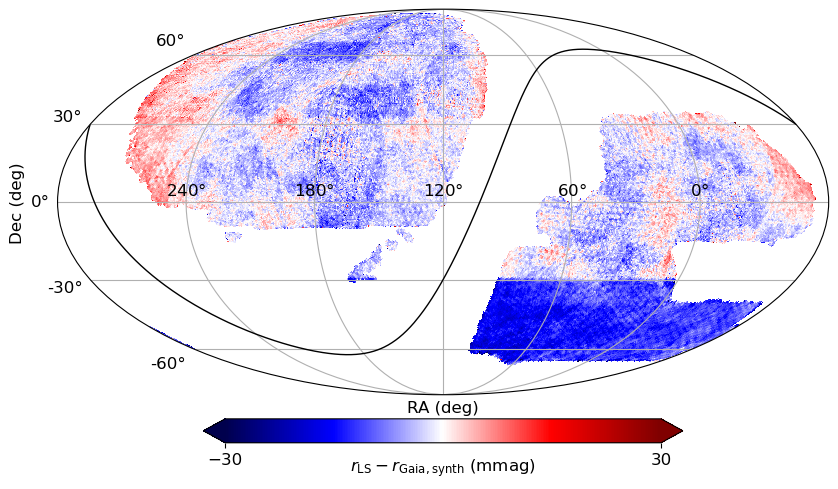}}
    \resizebox{0.6\columnwidth}{!}{\includegraphics{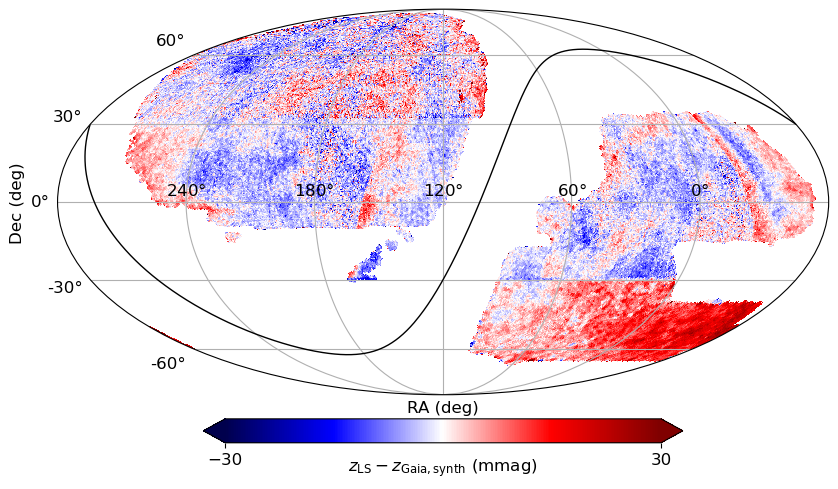}}
    \caption{Maps of the magnitude offset between LS DR9 and the Gaia synthetic photometry in $grz$ bands. The offset is the median value of the magnitude difference, $m_\mathrm{{LS}}-m_\mathrm{{Gaia,synth}}$, for stars in each HEALPix (Nside=128) pixel. We separate North and South at $\mathrm{DEC}=+32.375$ in the North Galactic Cap (i.e., only showing northern offsets at $\mathrm{DEC}>+32.375$). 
    The LS photometry is tied to PS1 north of $\mathrm{DEC}=-29.25^\circ$ and calibrated using ubercal south of this boundary.
    }
    \label{fig:dr9_offset_maps}
\end{figure}

To assess the spatial variation of the photometric offsets, we divide the cross-matched sample into HEALPix \citep{healpix} pixels with $\mathrm{Nside}=128$ and compute the median magnitude difference in each pixel. We remove pixels with $<20$ stars in each pixel (which removes 4\%-7\% of the pixels in DR9 South). Figure \ref{fig:dr9_offset_maps} shows the maps of the median magnitude difference in $grz$-bands between LS DR9 and Gaia synthetic photometry.
The relative calibration errors (i.e., the pixel-to-pixel \textsc{rms})  in $grz$ are 4.7, 3.7, 4.4~mmag in DR9 South; 4.6, 4.5, 5.9 mmag in DR9 North; and 5.0, 3.9, 4.3, 5.5 mmag in DR10 $griz$ bands. These estimates include calibration errors in both LS and Gaia, and represent the upper limit on the LS calibration error. The increase in DR10 calibration errors is due to its larger area coverage –– restricting to stars in both DR9 and DR10, their calibration \textsc{rms} agrees within 0.02 mmag. These \textsc{rms} calculations exclude the region at $\mathrm{Dec}<-29\degr$ (see below).

While photometric errors of individual stars contribute to the pixel-to-pixel \textsc{rms}, they are mostly negligible.
For smaller spatial pixels (i.e., $\mathrm{Nside}=256$) with $\geq20$ stars/pixel (removing 26-48\% pixels), we get relative calibration \textsc{rms} of 5.0, 3.9, 4.9 mmag in $grz$ in DR9 South. The higher \textsc{rms} compared to $\mathrm{Nside}=128$ is due to having fewer stars per pixel and thus larger per-pixel error, and to smaller-scale variations in the calibration systematics only probed at higher resolution.

A few patterns are visible:
1) Gaia scan patterns, which are most significant in the $g$ band.
2) Honeycomb patterns (especially in $r$ and $i$ bands), likely due to calibration errors in Pan-STARRS1 \citep{chambers_panstarrs1_2016} to which LS photometry is tied at $\mathrm{Dec}>-29.25\degr$.
3) Gradients at large angular scales, e.g., positive offsets towards the Galactic Plane at $\mathrm{RA}\simeq300\degr$.
It is unclear if these variations are due to calibration systematics in LS or Gaia or both.
4) Large overall offsets at $\mathrm{Dec}<-29.25\degr$ in $r$- and $z$-bands in DR9. At $\mathrm{Dec}=-29.25\degr$ the LS photometric calibration switches from PS1 to internal calibration\footnote{The transition happens at the CCD image level: CCD images centered at above $\mathrm{DEC}=-29.25\degr$ obtain their zero points from PS1, and from ubercal at below $\mathrm{DEC}=-29.25\degr$.} (the ``ubercalibration'' method; \citealt{padmanabhan_improved_2008, schlafly_photometric_2012}).
The $r$-band offset has a median value of -13 mmag with a slight gradient varying from -16 to -9 mmag. 
The $z$-band offset has a median value of 8 mmag with a stronger gradient varying from 1 to 19 mmag. 
These offsets are due to an absolute calibration error in DR9 (fixed in DR10). At $\mathrm{Dec}<-30\degr$, the DR9 calibration \textsc{rms} errors in $grz$ are 3.7, 3.3, 6.6 mmag; after removing the smoothly-varying components (the monopole and dipole), the $r$ and $z$ band calibration \textsc{rms} errors are reduced to 2.8 and 4.1 mmag.

\section{Discussion and summary} \label{sec:conclusion}

We have assessed the calibration precision of LS DR9 and DR10 photometry by comparing with Gaia synthetic photometry. Due to systematics in the Gaia synthetic photometry (which we cannot separate from the LS calibration systematics), we only obtain upper limits on the LS calibration error. We find spatial variations in the photometric offsets due to LS calibration systematics (i.e., the ``honeycomb'' patterns) as well as variations at larger angular scales. We also find large systematic offsets at $\mathrm{Dec}<-29.25\degr$ in $r$- and $z$-bands in DR9, which do not affect DESI as it only observes at $\mathrm{Dec}>-29.25\degr$. But for samples selected with the full DR9 imaging data, the offsets could cause a systematic shift in the surface density or in the photometric redshifts.

\begin{acknowledgments}
We thank Michele Bellazzini and Francesca De Angeli for useful suggestions and comments.
This research is supported by the Director, Office of Science, Office of High Energy Physics of the U.S. Department of Energy under Contract No. DE–AC02–05CH11231, and used the National Energy Research Scientific Computing Center, a DOE Office of Science User Facility under the same contract.
This research made use of data from the DESI Legacy Imaging Surveys (\url{https://www.legacysurvey.org/acknowledgment/} for the complete acknowledgments) and from the European Space Agency (ESA) mission
{\it Gaia} (\url{https://www.cosmos.esa.int/gaia}), processed by the {\it Gaia}
Data Processing and Analysis Consortium (DPAC,
\url{https://www.cosmos.esa.int/web/gaia/dpac/consortium}). Funding for the DPAC
has been provided by national institutions, in particular the institutions
participating in the {\it Gaia} Multilateral Agreement.  
\end{acknowledgments}


\bibliography{LS_vs_Gaia}{}
\bibliographystyle{aasjournal}

\end{document}